# Capacity Analysis for Continuous Alphabet Channels with Side Information, Part I: A General Framework

Majid Fozunbal, Steven W. McLaughlin, and Ronald W. Schafer*

School of Electrical and Computer Engineering
Georgia Institute of Technology
Atlanta, GA 30332-0250
{majid, swm, rws}@ece.gatech.edu

August 4, 2004

## Abstract

Capacity analysis for channels with side information at the receiver has been an active area of interest. This problem is well investigated for the case of finite alphabet channels. However, the results are not easily generalizable to the case of continuous alphabet channels due to analytic difficulties inherent with continuous alphabets.

In the first part of this two-part paper, we address an analytical framework for capacity analysis of continuous alphabet channels with side information at the receiver. For this purpose, we establish novel necessary and sufficient conditions for weak* continuity and strict concavity of the mutual information. These conditions are used in investigating the existence and uniqueness of the capacity-achieving measures. Furthermore, we derive necessary and sufficient conditions that characterize the capacity value and the capacity-achieving measure for continuous alphabet channels with side information at the receiver.

*This work was supported in part by Texas Instruments Leadership University Program.

# 1 Introduction

We consider the capacity analysis for continuous alphabet channels with side information at the receiver, i.e., channels where the input, output, state, and side information alphabets are abstract continuous spaces. For finite alphabet channels, this problem is well explored in the literature, e.g., [1], [2], [3], [4], [5], and [6]. However, the results for finite alphabet channels are not necessarily generalizable to continuous alphabet channels.

In fact, as shown by Csiszár [7], there are some technical difficulties that must be considered when working with continuous alphabet channels. Recall that in finite alphabet channels, the capacity analysis is performed over a finite dimensional space of input probability distributions, e.g., the simplex of input probability distributions. In this case, the mutual information is a real-valued function over the space of input distributions. As a result, the capacity analysis can be conducted over the Euclidean topology. Hence, one can simply verify the required global and local analytical properties of the set of input distributions and the mutual information. In contrast, for continuous alphabet channels, the capacity analysis needs to be conducted over the weak* topology. This requires completely different analytical tools and arguments that are based on machineries from measure theory and functional analysis.

In the first part of this two-part study, we introduce an analytical framework for capacity analysis of continuous alphabet channels with side information at the receiver. From the practical point of view, the results of this part are useful in capacity analysis for a large class of channels including fading channels with side information at the receiver. In these channels, since the channel state (realization) changes from time-to-time, new challenges are imposed in capacity analysis of the channel. Moreover, according to how much knowledge we have about the channel state ahead of the time, one might have a range of scenarios from no channel state information (CSI) to full CSI, see e.g., [8], [9], [10], [11], and [12]. Hence, a unified analytical framework is required that enables us to tackle the capacity analysis for different scenarios. In the first part of this paper, we address a general framework for capacity analysis of continuous alphabet channels followed by applications to the multiple antenna channels in the second part. Specifically, in this part, we address certain analytical properties of the space of input measures and the mutual information function based on notions from measure theory, functional analysis, and convex optimization.

The organization of this part is as follows. A brief introduction to the problem setup is given in Section 2. In Section 3, we introduce an analytical treatment of the space of input measures and the mutual information function and address issues such as the weak* compactness of the space of input measures along with strict concavity and weak* continuity of the mutual information. In Section 4, we raise the issue of capacity analysis and address necessary and sufficient conditions regarding the existence, uniqueness, and the expression of the capacity-achieving measure. Finally, Section 5 states some concluding remarks along with some guidelines for future work. A brief introduction to the required analytical preliminaries for this paper is given in Appendix A. A detailed investigation of applications of the results of this part to multiple antenna channels will be provided in the second part of this two-part paper.

## 2 Setup

In this section, we introduce the setup for continuous alphabet channels with side information. We assume a discrete-time memoryless channel (DTMC) where $X$, $Y$, $S$, and $V$ denote the input, output, state, and side information alphabets of a point-to-point communication channel. We assume that $X$, $Y$, $S$, and $V$ are *locally compact Hausdorff* (LCH) spaces [13], e.g., alphabets are like $\mathbb{R}^n$ (or $\mathbb{C}^n$) which are *separable* [14]. Moreover, the alphabets are assumed to be associated with a corresponding Borel $\sigma$-algebra; e.g., $(X, \mathcal{B}_X)$, $(Y, \mathcal{B}_Y)$, $(S, \mathcal{B}_S)$ are the Borel-measurable spaces denoting the input, output, and the state alphabets of DTMC, respectively; where $\mathcal{B}_X$, $\mathcal{B}_Y$, and $\mathcal{B}_S$ denote the Borel $\sigma$-algebras of $X$, $Y$, and $S$, respectively. The DTMC is represented by a collection of *Radon probability measures* [13] over $(Y, \mathcal{B}_Y)$ as follows,

$$\mathscr{W}_{X,S}(Y) = \{W(\cdot|x,s) \in \mathscr{P}(Y)|\ x \in X,\ s \in S\}, \tag{1}$$

where $\mathscr{P}(Y)$ is the collection of all Radon probability measures over $(Y, \mathcal{B}_Y)$. Note that the elements of the set $\mathscr{W}_{X,S}(Y)$ are probability measures over $(Y, \mathcal{B}_Y)$, that is, for each $x$ and $s$, $W(\cdot|x,s)$ is a probability measure on $(Y, \mathcal{B}_Y)$.

We assume that there exists some side information available at the receiver that is denoted by a measurable space $(V, \mathcal{B}_V)$ and characterized by a joint probability measure $Q \circ R$ over $Y \times V$. As a result, the side information is modelled by a conditional probability measure $Q_v$ over $(S, \mathcal{B}_S)$ for every $v \in V$. This is an appropriate model for side information, since it can model different scenarios. For example, one can observe that for the case of full channel state information (CSI), having $v$ there is no uncertainty on $S$, hence $Q_v$ is just the *dirac measure* [13]. On the other hand, when there exists no side information available at the receiver, the probability measure $Q_v$ is some measure $Q$ independent from $v$. As a result of existence of side information, the channel can be modelled by conditional probability measures on $(Y, \mathcal{B}_Y)$ as follows

$$\forall E \in \mathcal{B}_Y, \ W_{Q_v}(E|x) = \int W(E|x,s) dQ_v(s). \tag{2}$$

Having the above channel model, an *n-length block code* for the channel is a pair of mappings $(f, \phi)$ where $f$ maps some finite message set $\mathcal{M}$ into $X^n$ and $\phi$ maps $Y^n$ to $\mathcal{M}$. The mapping $f$ is called the *encoder* and the image of $\mathcal{M}$ under $f$ is called the *codebook*. Correspondingly, the mapping $\phi$ is called the *decoder* [1]. Assuming that the channel is memoryless, the channel from $X^n$ to $Y^n$ is governed by probability measures

$$\mathbf{W}_{\mathbf{Q_v}}^{(n)}(E_1 \times \cdots \times E_n|\mathbf{x}) = \prod_{i=1}^n W_{Q_{v_i}}(E_i|x_i),$$

which are conditional measures on the side information vector $\mathbf{v} = (v_1, v_2, \cdots, v_n) \in V^n$. Since the probability measure on $(V, \mathcal{B}_V)$ is $R$, then the average probability of error for transmission of message $m$ is defined by

$$e(m, \mathbf{W}^n, f, \phi) \triangleq 1 - \int \mathbf{W}_{\mathbf{Q_v}}^{(n)}(\phi^{-1}(m)|f(m)) dR^n,$$

and the maximum probability of error is defined by $e(\mathbf{W}^n, f, \phi) \triangleq \max_m e(m, \mathbf{W}^n, f, \phi)$. The channel coding problem is to make the message set $\mathcal{M}$ (the rate) as large as possible while keeping the maximum probability of error arbitrarily low, subject to some constraints applied to the choice of codebook.

A non-negative rate $\mathsf{R}$ for the channel is an $\epsilon$-achievable rate, if for every $\delta > 0$ and every sufficiently large $n$ there exist $n$-length codes of rate exceeding $\mathsf{R} - \delta$ and probability of error less than $\epsilon$. Correspondingly, the rate $\mathsf{R}$ is an achievable rate if it is $\epsilon$-achievable for all $0 < \epsilon < 1$. The supremum of achievable rates is called the *channel capacity*.

There are a number of problems that need to be addressed in capacity assessment of a channel: These include the capacity value and the existence, the uniqueness, and the characterization of the capacity-achieving input measures. In this part of this two-part study, we introduce a framework to address the above problems in a unified manner for different classes of channels.

## 3  An analytical treatment

In capacity analysis of communication channels, there are often some constraints applied to the transmitted signals. Commonly, this is in the form of a maximum or an average energy constraint [15]. A maximum energy constraint is translated into a restriction of the input alphabet to a bounded subset of $X$.[1] On the other hand, an average energy constraint is translated to input measures with a second moment constraint. Restriction of input probability measures by higher moment constraints or a combination of moment constraints and a bounded alphabet are also considered in practice, see e.g., [15], [16]. Since the capacity analysis problem is a convex optimization problem, it is of interest to know whether such a restricted collection of input probability measures is convex and compact (in a certain sense). Moreover, since we try to optimize the mutual information over such a collection, we need to investigate the global and local analytical behavior of the mutual information over the space of input measures.

In this section, we address some analytical notions and properties of the space of input measures and the mutual information that are essential to the capacity analysis of continuous alphabet channels. We assume that a reader has elementary background in functional analysis. However, a reader can refer to Appendix A to grasp a general view of the analytical preliminaries that are used throughout the paper. For the sake of conciseness, we only express the main results in this section and we address the details in Appendix A.

---

[1] For example, applications that use a hard-limiter power amplifier.

## 3.1 Weak* compactness of the space of input probability measures

Let $(X, \mathcal{B}_X)$ be an LCH Borel-measurable space. Let $\mathscr{M}(X)$ denote the space of Radon measures over $(X, \mathcal{B}_X)$. In probability theory, where the objects of interest are the set of probability measures $\mathscr{P}(X) \subset \mathscr{M}(X)$, *weak* topology*, the weakest topology over $\mathscr{M}(X)$, is used to investigate the analytical properties of the functionals that are defined over $\mathscr{P}(X)$.

In weak* topology, the convergence phenomenon is called *weak* convergence*[2] and defined as follows. A sequence of probability measures converges weakly*, denoted by $P_n \xrightarrow{w^*} P$ if and only if $\int f dP_n \to \int f dP$ for all $f \in C_b(X)$, where

$$C_b(X) = \{f : X \to \mathbb{R} | \ f \text{ is continuous and bounded}\}$$

denotes the set of all bounded continuous functions.

Corresponding to the definition of weak* convergence, we have a notion of compactness which is called *weak* compactness*. That is, a family of probability measures $\mathscr{P}_A(X) \subseteq \mathscr{P}(X)$ is *relatively weak* compact* if every sequence of measures in $\mathscr{P}_A(X)$ contains a subsequence which converges weakly* (see Appendix A) to a probability measure in the closure of $\mathscr{P}_A(X)$.[3] In general, verification of relative compactness of probability measures over an abstract space is not an easy task. However, for *complete, separable* spaces [13], there is a simple way to verify this property, as follows.

A family of probability measures $\mathscr{P}_A(X) \subset \mathscr{P}(X)$ is *tight* if for every $\varepsilon > 0$, there is a compact set $K \subseteq X$ such that $\sup_{P \in \mathscr{P}_A(X)} P(K^c) \leq \varepsilon$. Based on this definition, we restate Prokhorov's Theorem from [17].

**Theorem 3.1 (Prokhorov's Theorem).** *Let $\mathscr{P}_A(X)$ be a family of probability measures defined over the complete separable measurable space $(X, \mathcal{B}_X)$. Then $\mathscr{P}_A(X)$ is relatively weak* compact if and only if it is tight.*

*Proof.* See [17, p. 318] □

As a result, for $X = \mathbb{R}^n$(or $\mathbb{C}^n$) together with the Borel $\sigma$-algebra $\mathcal{B}_X$, it suffices only to check the hypothesis of Prokhorov's Theorem. Using Prokhorov's Theorem, [7] derived the following sufficient condition for compactness of a restricted collection of probability measures.

---

[2] In textbooks on probability theory, the term *vague* is used instead of weak*.
[3] Note that the term "relative" refers to the compactness of closure.

**Lemma 3.1.** *Let $g : X \to \mathbb{R}^k$ be a nonnegative Borel-measurable function such that the set $K_L = \{x \in X | g_i(x) \leq L_i, \ i = 1, \cdots k\}$ is compact for every $L \in \mathbb{R}^{+k}$. Then, the collection*

$$\mathscr{P}_{g,\Gamma}(X) = \left\{ P \in \mathscr{P}(X) \middle| \int g_i(x) dP \leq \Gamma_i, \ i = 1, \cdots k \right\},$$

*is tight and closed, and hence weak\* compact for every $\Gamma \in \mathbb{R}^{+k}$.*

*Proof.* See [7, Lem. 1]. □

Note that Lemma 3.1 holds in general for a collection of constraints defined by positive functions $\{g_i\}$ and positive values $\{\Gamma_i\}$ such that each $g_i$ satisfies the hypothesis of Lemma 3.1. As an example of the usage of Lemma 3.1, one can consider $X = \mathbb{R}^n$ along with a restricting function $g(x) = \|x\|_2^2$ and a fixed positive value $\Gamma > 0$ to easily verify that the set of probability measures with a second moment constraint, $\mathscr{P}_{g,\Gamma}(X)$, is compact. Likewise, if $A$ is a compact subset of $X$, one can consider

$$g(x) = \begin{cases} \|x\|_2^2, & x \in A \\ +\infty, & \text{otherwise} \end{cases}.$$

and a fixed positive value $\Gamma > 0$ to easily verify that $\mathscr{P}_{g,\Gamma}(X)$ is compact.

## 3.2 Mutual information

In this subsection, we provide conditions for *weak\* continuity* (see Appendix A) of the mutual information over a set of probability measure. We also state and prove some novel conditions for strict concavity of the mutual information. Applications of these properties will be explored in the next section, where they will be used to address the existence, the uniqueness, and the characterization of the capacity-achieving measure for continuous alphabet channels with side information.

### 3.2.1 Definition

To present the precise expression of the mutual information, following [7] and [18], we first express the definition of *informational divergence or relative entropy* as follows.

For a given measurable space $(X, \mathcal{B}_X)$, consider two probability measures $P$ and $Q$. The informational divergence between these two measures is [18] defined by

$$D(P\|Q) \triangleq \sup\left\{\sum_{i=1}^{N} P(E_i) \log_2 \frac{P(E_i)}{Q(E_i)} : N \in \mathbb{N},\ E_i \in \mathcal{B}_X \text{ disjoint, and } X = \bigcup_{i=1}^{N} E_i\right\}. \quad (3)$$

This can be viewed as the generalization of relative entropy of probability measures of finite sets to the probability measures of infinite sets. By (3), it appears that if there exists an $E_i \in \mathcal{B}_X$ such that $Q(E_i) = 0$ but $P(E_i) \neq 0$, then $D(P\|Q) = \infty$. Thus, a necessary condition to have a finite relative entropy between $P$ and $Q$ is that for every $E \in \mathcal{B}_X$ with $Q(E) = 0$, $P(E) = 0$. But this means that $P$ is *absolutely continuous* with respect to $Q$ denoted by $P \ll Q$ (see Appendix A).

By the log-sum inequality [19], it can be verified that for each partition in the right-hand side (RHS) of (3), consequent refined partitioning increases the value of the summation. In fact, as the partitions get finer, the finite sum in the RHS of (3) gets closer to $D(P\|Q)$. This observation provides intuition into an important result of [18] which expresses $D(P\|Q)$ as

$$D(P\|Q) = \begin{cases} \int \log_2 \frac{dP}{dQ} dP, & \text{if } P \ll Q \\ +\infty, & \text{otherwise,} \end{cases} \quad (4)$$

where $\frac{dP}{dQ}$ is the density of $P$ with respect to $Q$ [13, p. 91]. In fact, the condition $P \ll Q$ is a necessary and sufficient condition for the finiteness of the informational divergence as we show below.

**Proposition 3.1.** *For a pair of probability measures $P$ and $Q$, $D(P\|Q) < \infty$ if and only if $P \ll Q$. Furthermore, $\int |\log_2 \frac{dP}{dQ}| dP < \infty$ if and only if $P \ll Q$.*

*Proof.* The direct part of this statement is proved in [18, p. 20] which is observed by (4). Suppose the inverse part is not true. That is $P \ll Q$, but $\int \log_2 \frac{dP}{dQ} dP = \infty$. Because $P$ is a finite measure, then for the set $E = \{x \in X : \frac{dP}{dQ} = \infty\}$ we must have $P(E) > 0$. On the other hand, since $P(E) = \int_E \frac{dP}{dQ} dQ$, this requires that $Q(E) = 0$. This is a contradiction to the hypothesis that $P \ll Q$. Using the inequality $\int |\log_2 \frac{dP}{dQ}| dP \leq D(P\|Q) + \frac{2}{e \ln 2}$ from [18, p. 20], we conclude the rest of the proof. □

By the Lebesgue-Radon-Nickodym Theorem [13, p. 90], there exists a positive real valued function $f = \frac{dP}{dQ}$ such that $P(E) = \int_E f dQ$. Thus, for $P \ll Q$,

$$D(P\|Q) = \int \frac{dP}{dQ} \log_2 \frac{dP}{dQ} dQ = \int f \log_2 f dQ. \tag{5}$$

Using the expression of relative entropy as (4) and (5), we now introduce a precise expression of mutual information function.

Let $(X, \mathcal{B}_X)$ bs the input and $(Y, \mathcal{B}_Y)$ be the output measurable spaces of a channel. The product space of $X$ and $Y$ is denoted by $(X \times Y, \mathcal{B}_X \otimes \mathcal{B}_Y)$, where $\mathcal{B}_X \otimes \mathcal{B}_Y$ is the Borel $\sigma$-algebra induced on $X \times Y$. Let $P$ and $T$ be two probability measures over them, respectively. The probability measure that is induced on $(X \times Y, \mathcal{B}_X \otimes \mathcal{B}_Y)$ is denoted by $P \times T$ which is defined as follows,

$$\forall E \in \mathcal{B}_x \otimes \mathcal{B}_Y, \ (P \times T)(E) = \iint_E d(P \times T) = \int \int_{E_y} dP dT$$

where for every $y \in Y$, $E_y = \{x \in X | (x,y) \in E\}$.

As mentioned before, since side information is available at the receiver, the channel is described by probability measures $W_{Q_v}(\cdot|x)$ defined as in (2). For an input probability measure $P$, let the joint conditional measure of the input and output denoted by $P \circ W_{Q_v}$ and let the marginal output measure denoted by $PW_{Q_v}$. defined as follows. For every $A \times B \in \mathcal{B}_X \times \mathcal{B}_Y$, we have

$$P \circ W_{Q_v}(A \times B) = \int_A W_{Q_v}(B|x) dP,$$

which results into a marginal probability measure on $(Y, \mathcal{B}_Y)$ such that,

$$PW_{Q_v}(B) = \int W_{Q_v}(B|x) dP.$$

It can be verified that $P \circ W_{Q_v} \ll P \times PW_{Q_v}$. On the other hand, $P \circ W_{Q_v} \ll P \times PW_{Q_v}$ if and only if $W_{Q_v}(\cdot|x) \ll PW_{Q_v}$ $P$-a.e. As a result, following [7], we can express the mutual information of the channel as

$$I(P, W_{Q_v}|R) = \iint D(W_{Q_v}(\cdot|x) \| PW_{Q_v}) dP dR$$
$$= \iiint \log_2 \frac{dW_{Q_v}(\cdot|x)}{d(PW_{Q_v})} dW_{Q_v}(\cdot|x) dP dR \tag{6}$$

where $R$ denotes the probability measure on the space of channel state information $(V, \mathcal{B}_V)$. To emphasize that the mutual information is a function over $\mathscr{P}_{g,\Gamma}(X)$, we deliberately use a different notation for it (as in [1]) rather than the more common notation expressed in terms of random variables [19]. In the following subsections, we investigate some global and local analytical properties of the mutual information (6) such as concavity and continuity.

### 3.2.2 Convexity and concavity

In this part, we address some global analytical properties of the mutual information. For this purpose, we first study these global properties of the relative entropy and then we generalize them for mutual information.

The convexity of relative entropy with respect to the convex combination of a pair of measures is well known [18]. However, to the best of our knowledge, necessary and sufficient conditions for its strict convexity were not known before. This is of particular interest to show the uniqueness of the capacity-achieving measure, as it will be shown later. Hence, in the following theorem, we state necessary and sufficient conditions for the strict convexity of relative entropy.

**Theorem 3.2.** $D(P\|Q)$ *is convex with respect to the pair* $(P,Q)$. *That is, for given pairs* $(P_1, Q_1)$ *and* $(P_2, Q_2)$ *and given scalar* $0 < \alpha < 1$,

$$D(\alpha P_1 + (1-\alpha)P_2 \| \alpha Q_1 + (1-\alpha)Q_2) \leq \alpha D(P_1\|Q_1) + (1-\alpha)D(P_2\|Q_2).$$

*Moreover, the inequality is strict if and only if there exists a set* $E \in \mathcal{B}_X$ *such that* $\frac{dP_1}{dQ_1} \neq \frac{dP_2}{dQ_2} \neq 0$ *on* $E$ *and for all nonempty Borel-measurable* $F \subseteq E, F \in \mathcal{B}_X, P_1(F) \neq 0$ *and* $P_2(F) \neq 0$.

*Proof.* For convenience in derivations, let $\beta = 1 - \alpha$. Then, it can be verified that $Q_1 \ll \alpha Q_1 + \beta Q_2$ and $Q_2 \ll \alpha Q_1 + \beta Q_2$. Let $g_1$ and $g_2$ denote the density functions of $Q_1$ and $Q_2$ with respect to $\alpha Q_1 + \beta Q_2$, respectively. That is $dQ_1 = g_1 d(\alpha Q_1 + \beta Q_2)$ and $dQ_2 = g_2 d(\alpha Q_1 + \beta Q_2)$. Note that $\alpha g_1 + \beta g_2 = 1$. Since $P_1 \ll Q_1$ and $P_2 \ll Q_2$ associated with density functions $f_1 = \frac{dP_1}{dQ_1}$ and $f_2 = \frac{dP_1}{dQ_2}$, then $\alpha P_1 + \beta P_2 \ll \alpha Q_1 + \beta Q_2$ and $d(\alpha P_1 + \beta P_2) = (\alpha f_1 g_1 + \beta f_2 g_2) d(\alpha Q_1 + \beta Q_2)$. Thus,

$$D(\alpha P_1 + \beta P_2 \| \alpha Q_1 + \beta Q_2) = \int (\alpha f_1 g_1 + \beta f_2 g_2) \log_2(\alpha f_1 g_1 + \beta f_2 g_2) d(\alpha Q_1 + \beta Q_2)$$

$$\leq \alpha \int f_1 \log_2 f_1 dQ_1 + \beta \int f_2 \log_2 f_2 dQ_2 \text{ (Log-sum inequality)}$$

$$= \alpha D(P_1\|Q_1) + \beta D(P_2\|Q_2).$$

For strictness of the inequality, note that in log-sum inequality, for $x \in X$, strict inequality occurs if $f_1(x) \neq f_2(x) \neq 0$ and $g_1(x) \neq 0, g_2(x) \neq 0$. Let $N$ denote the maximal null set of $\alpha Q_1 + \beta Q_2$, then define

$$E = \{x \in X \backslash N : f_1(x) \neq f_2(x) \neq 0, \ g_1(x) \neq 0, \ g_2(x) \neq 0\}.$$

This set is Borel measurable since $f_1, f_2, g_1, g_2$ are Borel measurable. To have strict inequality, we need $E$ such that $(\alpha Q_1 + \beta Q_2)(E) \neq 0$. Because, $g_1$ and $g_2$ are nonzero over $E$, $Q_1(E) \neq 0$ and $Q_2(E) \neq 0$. Since $f_1$ and $f_2$ are also nonzero, $P_1(E) \neq 0$ and $P_2(E) \neq 0$. For every nonempty Borel-measurable subset $F \subseteq E$, the above argument holds. This proves the direct part of the assertion.

On the other hand, suppose there exists $E \in \mathcal{B}_X$ with the above definitions such that for every nonempty Borel-measurable $F \subseteq E$, $P_1(F) \neq 0$, $P_2(F) \neq 0$, and $f_1 \neq f_2$ over $F$. Let $K_i = \{x \in X \backslash N : g_i(x) \neq 0\}$ for $i = 1, 2$. It is clear that both $E \cap K_1 \neq \emptyset$ and $E \cap K_2 \neq \emptyset$, otherwise either $P_1(E) = 0$ or $P_2(E) = 0$ which is a contradiction to our hypothesis. This means that $(E \cap K_i) \subset E$ is a proper subset of $E$, and by hypothesis, $P_i(E \cap K_j) \neq 0$ ($i, j \in \{1, 2\}$). This implies that $(E \cap K_1) \cap (E \cap K_2) \neq \emptyset$. By definition of $(E \cap K_1) \cap (E \cap K_2)$, we deduce that $(\alpha Q_1 + \beta Q_2)(E \cap K_1 \cap K_2) \neq 0$. Thus for the set $E \cap K_1 \cap K_2$ log-sum inequality holds strictly. Hence, the inequality would be strict. This concludes the proof. $\square$

As an special case of Theorem 3.2, we obtain the following corollary.

**Corollary 3.1.** *If $Q = Q_1 = Q_2$ in Theorem 3.2, then the convexity is strict if and only if there exists a set*

$$K = \left\{x \in X : \frac{dP_1}{dQ} \neq \frac{dP_2}{dQ} \neq 0\right\}$$

*such that $Q(K) > 0$.*

*Proof.* From Theorem 3.2, the strict inequality holds if and only if there exists $E \in \mathcal{B}_X$ such that $\frac{dP_1}{dQ} \neq \frac{dP_2}{dQ} \neq 0$ on $E$ and for every proper $F \subset E \in \mathcal{B}_X$, $P_1(F) > 0$ and $P_2(F) > 0$. Taking a nonempty $K \subseteq E$, the direct part of the assertion is proved.

For the reverse part, suppose there exists a set $K$ as in the hypothesis. Let $N$ be the maximal null set of $Q$ and let $E = K \backslash N$. Now, it can be verified that for any proper

Borel-measurable subset $F \subset E$, we have $P_1(F) > 0$ and $P_2(F) > 0$. This proves the reverse direction of the assertion. □

Now, we use Theorem 3.2 and Corollary 3.1 to establish a proposition on global properties of the mutual information. This can be considered as a generalization of a similar result in [7] for channels with side information. However, we provide a rigorous proof for this more general proposition, since later in the paper, we use some of the intermediate results.

**Proposition 3.2.** *The mutual information* (6) *is concave with respect to P, convex with respect to* $W_{Q_v}(\cdot|x)$, *and linear with respect to R.*[4]

*Proof.* The linearity with respect to $R$ is clearly seen by (6). The convexity with respect to $W_{Q_v}(\cdot|x)$ follows by the convexity of $D(W_{Q_v}(\cdot|x)\|PW_{Q_v})$ which can be verified by Theorem 3.2.

To prove the concavity with respect to the input distribution $P$, let $0 < \alpha < 1$, $\beta = 1-\alpha$, and $P = \alpha P_1 + \beta P_2$. By linearity, this implies that $PW_{Q_v} = \alpha P_1 W_{Q_v} + \beta P_2 W_{Q_v}$. Pick an auxiliary probability measures $T_v$ (conditional on $v$) over $Y$ such that $PW_{Q_v} \ll T_v$; the existence of such a measure is obvious. Since, $W_{Q_v}(\cdot|x) \ll PW_{Q_v}$ and $PW_{Q_v} \ll T_v$, then $W_{Q_v}(\cdot|x) \ll T_v$. By Proposition 3.1, we also know that $D(PW_{Q_v}\|T_v) < \infty$. Let us consider the mutual information for a fixed value $v$, and denote it by $I(P, W_{Q_v})$. As a result, we can expand it as

$$I(P, W_{Q_v}) = \iint \log_2 \frac{dW_{Q_v}(\cdot|x)}{d(PW_{Q_v})} dW_{Q_v}(\cdot|x) dP$$
$$= \iint \left[\log_2 \frac{dW_{Q_v}(\cdot|x)}{dT_v} - \log_2 \frac{d(PW_{Q_v})}{dT_v}\right] dW_{Q_v}(\cdot|x) dP$$
$$= \iint \log_2 \frac{dW_{Q_v}(\cdot|x)}{dT_v} dW_{Q_v}(\cdot|x) dP - \iint \log_2 \frac{d(PW_{Q_v})}{dT_v} dW_{Q_v}(\cdot|x) dP$$

Now, we can use Fubini's theorem to change the order of integration in the second term and

---
[4]Note that concavity, convexity, and linearity are with respect to the convex combination of the operands.

apply Theorem 3.2 to obtain:

$$I(P, W_{Q_v}) = \iint \log_2 \frac{dW_{Q_v}(\cdot|x)}{dT_v} dW_{Q_v}(\cdot|x) dP - \int \log_2 \frac{d(PW_{Q_v})}{dT_v} d(PW_{Q_v}) \quad (7)$$
$$\geq \alpha \iint \log_2 \frac{dW_{Q_v}(\cdot|x)}{dT_v} dW_{Q_v}(\cdot|x) dP_1 - \alpha \int \log_2 \frac{d(P_1 W_{Q_v})}{dT_v} d(P_1 W_{Q_v})$$
$$+ \beta \iint \log_2 \frac{dW_{Q_v}(\cdot|x)}{dT_v} dW_{Q_v}(\cdot|x) dP_2 - \beta \int \log_2 \frac{d(P_2 W_{Q_v})}{dT_v} d(P_2 W_{Q_v})$$

Noting that $P_1 W_{Q_v} \ll T_v$, $P_2 W_{Q_v} \ll T_v$ and using the above arguments, we can contract the RHS to obtain

$$I(P, W_{Q_v}) \geq \alpha I(P_1, W_{Q_v}) + \beta I(P_2, W_{Q_v}).$$

Because, this holds for every $v$, we can integrate both sides of the above equation with respect to $R$ and deduce that

$$I(P, W_{Q_v}|R) \geq \alpha I(P_1, W_{Q_v}|R) + \beta I(P_2, W_{Q_v}|R).$$

This concludes the proof. $\square$

Proposition 3.2 addresses the concavity of the mutual information with respect to input measures. In the following proposition, we address its strictness.

**Proposition 3.3 (Strictness).** *The mutual information is strictly concave with respect to the input measure if and only if the set*

$$E = \left\{ (y, v) \in Y \times V : \frac{d(P_1 W_{Q_v})}{d(PW_{Q_v})} \neq \frac{d(P_2 W_{Q_v})}{d(PW_{Q_v})} \neq 0 \right\}$$

*has $(PW_{Q_v} \times R)(E) > 0$. Moreover, if $T_v$ is a conditional probability measure on $Y$ such that $PW_{Q_v} \ll T_v$ for all $v \in V$, then strict concavity holds if and only if the set*

$$E = \left\{ (y, v) \in Y \times V : \frac{d(P_1 W_{Q_v})}{dT_v} \neq \frac{d(P_2 W_{Q_v})}{dT_v} \neq 0 \right\}$$

*has nonzero measure with respect to the product measure $T_v \times R$.*

*Proof.* The proof follows from considering the proof of Proposition 3.2 together with Corollary 3.1. For a fixed $v$, by Corollary 3.1 if there exists a set $E_v$ such that $\frac{d(P_1 W_{Q_v})}{d(PW_{Q_v})} \neq \frac{d(P_2 W_{Q_v})}{d(PW_{Q_v})} \neq 0$ and $PW_{Q_v}(E_v) > 0$, then strictness holds. To have strictness in total, we need to have it for $R$-almost everywhere. The proof of the special case is immediate by definition of $E$. $\square$

This concludes our discussion on convexity and concavity properties of the mutual information.

### 3.2.3 Continuity

So far, we have discussed the compactness of the set of input probability measures and some global properties of the mutual information. In this subsection, we discuss the continuity of the mutual information in the sense of weak* topology. However, before expressing the main result of this part, let us introduce a useful inequality.

**Lemma 3.2.** *For a channel with side information as specified by $W_{Q_v}(\cdot|x)$ (2), let*

$$|I|(P, W_{Q_v}|R) \triangleq \iiint \left| \log_2 \frac{dW_{Q_v}(\cdot|x)}{d(PW_{Q_v})} \right| dW_{Q_v}(\cdot|x) dP dR.$$

*Then, the following inequalities hold*

$$I(P, W_{Q_v}|R) \leq |I|(P, W_{Q_v}|R) \leq I(P, W_{Q_v}|R) + \frac{2}{e \ln 2}.$$

*Proof.* The first inequality is obvious. The second inequality follows from a simple observation that $-\frac{1}{e \ln 2} \leq x \log_2 x$. As a result, we have $|x \log_2 x| \leq x \log_2 x + \frac{2}{e \ln 2}$. Using this observation, the proof of the second inequality follows. □

We now state and prove a novel sufficient condition for the continuity of mutual information.

**Theorem 3.3.** *Consider a channel with side information which is described by $W_{Q_v}(\cdot|x)$, together with a closed collection of input probability measures $\mathscr{P}_A(X)$. Suppose there exists a measure $T$ on $(Y, \mathcal{B}_Y)$ such that $W_{Q_v}(\cdot|x) \ll T$ and density function $f_{T,Q_v}(y|x) \triangleq \frac{dW_{Q_v}(\cdot|x)}{dT}$. If*

 a. *The function $f_{T,Q_v}(y|x)$ is continuous over $X \times Y \times V$, and $f_{T,Q_v}(y|x) \log_2 f_{T,Q_v}(y|x)$ is uniformly integrable over $\{T \times P \times R \,|\, P \in \mathscr{P}_A(X)\}$.*

 b. *For fixed $y$ and $v$, the function $f_{T,Q_v}(y|x)$ is uniformly integrable over $\mathscr{P}_A(X)$.*

*Then, the mutual information function is bounded and weak* continuous over $\mathscr{P}_A(X)$.*

*Proof.* To show the continuity of $I(P, W_{Q_v}|R)$, we need to show that for every sequence $P_n \xrightarrow{w^*} P$, we have $I(P_n, W_{Q_v}|R) \to I(P, W_{Q_v}|R)$. For this purpose, using Proposition 3.1, similar to the proof of Proposition 3.2, we decompose the conditional mutual information into two terms.

$$I(P, W_{Q_v}|R) = \iiint \log_2 \frac{dW_{Q_v}(\cdot|x)}{d(PW_{Q_v})} dW_{Q_v}(\cdot|x) dP dR$$
$$= \iiint \log_2 \frac{dW_{Q_v}(\cdot|x)}{dT} dW_{Q_v}(\cdot|x) dP dR - \iint \log_2 \frac{d(PW_{Q_v})}{dT} d(PW_{Q_v}) dR$$
$$= \iiint f_{T,Q_v}(y|x) \log_2 f_{T,Q_v}(y|x) dT dP dR - \iint f_{T,P,Q_v}(y) \log_2 f_{T,P,Q_v}(y) dT dR.$$

Momentarily, we assume that both terms are finite, then we provide evidence for this assumption. Thus, we need only to show that both terms are bounded and continuous over $\mathscr{P}_A(X)$.

*Continuity of the first term:* Since $P_n \xrightarrow{w^*} P$, by Proposition A.2, we have $T \times P_n \times R \xrightarrow{w^*} T \times P \times R$. Because $f_{T,Q_v}(y|x)$ is continuous, so is $f_{T,Q_v}(y|x) \log_2 f_{T,Q_v}(y|x)$. By hypothesis, $f_{T,Q_v}(y|x) \log_2 f_{T,Q_v}(y|x)$ is uniformly integrable over $\{T \times P \times R \,|\, P \in \mathscr{P}_A(X)\}$ (Definition A.2). Therefore, using Theorem A.2, we deduce that

$$\iiint f_{T,Q_v}(y|x) \log_2 f_{T,Q_v}(y|x) dT dP_n dR \to \iiint f_{T,Q_v}(y|x) \log_2 f_{T,Q_v}(y|x) dT dP dR.$$

This proves the continuity of the first term. The finiteness of the first term is immediate by the uniform integrability property.

*Continuity of the second term:* For fixed $y$ and $v$, since $f_{T,Q_v}(y|x)$ is uniformly integrable over $\mathscr{P}_A(X)$, by Theorem A.2, we deduce that $P_n \xrightarrow{w^*} P$ implies the pointwise convergence of $f_{T,P_n,Q_v}(y) \to f_{T,P,Q_v}(y)$. By continuity of the $\log_2$, we deduce the pointwise convergence of $f_{T,P_n,Q_v}(y) \log_2 f_{T,P_n,Q_v}(y) \to f_{T,P,Q_v}(y) \log_2 f_{T,P,Q_v}(y)$. It only remains to show the convergence of their integrals with respect to $T \times R$. For this purpose, we proceed as follows.

By Lemma 3.2 and its proof along with the log-sum inequality, for every $n$,

$$|f_{T,P_n,Q_v}(y) \log_2 f_{T,P_n,Q_v}(y)| \leq \frac{2}{e \ln 2} + f_{T,P_n,Q_v}(y) \log_2 f_{T,P_n,Q_v}(y)$$
$$\leq \frac{2}{e \ln 2} + \int f_{T,Q_v}(y|x) \log_2 f_{T,Q_v}(y|x) dP_n.$$

But, we have already shown that the integration of the RHS over $T \times R$ leads to a convergent sequence of integrals. Thus, by the generalized Dominated Convergence Theorem [13, p. 59],

we deduce that
$$\iint f_{T,P_n,Q_v}(y)\log_2 f_{T,P_n,Q_v}(y)dTdR \to \iint f_{T,P,Q_v}(y)\log_2 f_{T,P,Q_v}(y)dTdR$$
This implies the continuity of the second term. Note that its finiteness is obvious. Since both terms are finite and continuous, we deduce the continuity of mutual information. This concludes the proof. □

So far, we have discussed the conditions for compactness of the set of input probability measures and the strict concavity and continuity of the mutual information. The following section demonstrates the application of these results for capacity analysis purposes.

## 4 Capacity analysis

In this section, we address the capacity analysis for continuous alphabet channels with side information. We provide a coding and converse coding argument for the capacity value of the channels of our interest, and we address the existence, the uniqueness, and the characterization of the capacity-achieving input measure.

### 4.1 Channel capacity

Consider the channel of interest described by $W_{Q_v}(\cdot|x)$. Let $g: X \to \mathbb{R}^k$ be a nonnegative Borel-measurable function that satisfies the hypothesis of Lemma 3.1. Let $\Gamma \in \mathbb{R}^{+k}$ and $\mathscr{P}_{g,\Gamma}(X)$ defined as in Lemma 3.1. We show that

$$C = \sup_{P \in \mathscr{P}_{g,\Gamma}(X)} I(P, W_{Q_v}|R) \tag{8}$$

is the capacity of the channel. For this purpose, we use the results of [7] to express the coding and converse coding theorem for the case of continuous alphabet channels with side information.

**Lemma 4.1 (Converse Coding Lemma).** *Consider a collection of probability measures $\mathscr{P}_{g,\Gamma}(X)$ on $X$. For any $\delta > 0$, there exists $n_0$ and $\epsilon > 0$ such that for every code $(f, \phi)$ of length $n \geq n_0$ with $N$ codewords whose empirical measures all belong to $\mathscr{P}_{g,\Gamma}(X)$, if*

$$\frac{1}{n}\log_2 N > \sup_{P \in \mathscr{P}_{g,\Gamma}(X)} I(P, W_{Q_v}|R) + \delta,$$

*then the maximum error probability satisfies $e(\mathbf{W}^n, f, \phi) > \epsilon$.*

*Proof.* The proof follows from [7, Lemma 6]. We note that, here, the channel has only one strategy and we can consider $Y \times V$ as the output alphabet of our channel. Since $V$ is independent of the input, we can simplify the results of [7, Lemma 6] to obtain our assertion. $\square$

By Lemma 4.1, we can easily verify that any rate $R > C$ is not achievable. Suppose not, i.e., suppose $R > C$ is achievable. That is for every $\delta > 0$ and $\epsilon > 0$ there exists $n_0$ such that for every $n > n_0$, there exists a code with at least $\lceil 2^{n(R-\delta)} \rceil$ and error probability less than $\epsilon$. But this is a contradiction to the assertion of Lemma 4.1.

Now, inspired by [7, Thm. 1], we state the coding theorem.

**Theorem 4.1 (Coding Theorem).** *For every positive number $\delta$, there exists an integer $n_0$ and $\gamma > 0$ such that for block length $n \geq n_0$ for any prescribed codeword type $P \in \mathscr{P}_{g,\Gamma}(X)$ there exists a code with $N$ codewords, each of type $P$, such that*

$$\frac{1}{n} \log_2 N > I(P, W_{Q_v}|R) - \delta \ \ and \ \ e(\mathbf{W}^n, f, \phi) < 2^{-n\gamma}.$$

*Proof.* The proof is by [7, Thm. 1]. First consider $Y \times V$ as the output alphabet of the channel. Then, noting that the CSI, $V$, is independent from the input, we can simplify the results of [7, Thm. 1] to obtain our assertion. $\square$

Since the result of Theorem 4.1 holds for every input measure, it holds for their supremum. Hence, we can deduce that for every $\delta > 0$ and sufficient large block length, there exist codes with rate $R > \sup_{P \in \mathscr{P}_{g,\Gamma}(X)} I(P, W_{Q_v}|R) - \delta$. Because this is true for every $\delta > 0$, using the Converse Coding Lemma, we deduce that the channel capacity is

$$C = \sup_{P \in \mathscr{P}_{g,\Gamma}(X)} I(P, W_{Q_v}|R).$$

## 4.2 Existence

In this subsection, we give a sufficient condition for the existence of an optimal input measure, say $P_o$, such that the capacity is achievable by some code with codewords of type $P_o$.

**Proposition 4.1.** *Let $\mathscr{P}_A(X)$ denote a weak\* compact collection of probability measures on $(X, \mathcal{B}_X)$ and let the channel be described by $W_{Q_v}(\cdot|x)$. If the mutual information $I(P, W_{Q_v}|R)$ is continuous over $\mathscr{P}_A(X)$, then it is bounded and achieves its maximum on $\mathscr{P}_A(X)$.*

*Proof.* We claim that the range of $I(P, W_{Q_v}|R)$ is bounded. Suppose not. Then, for every $n \in N$, there exists $P_n \in \mathscr{P}_A(X)$ such that $I(P_n, W_{Q_v}|R) \geq n$. But the sequence $(P_n)_{n=1}^\infty$ belongs to $\mathscr{P}_A(X)$ which is a weak\* compact family. By definition this means that there exists a weak\* convergent subsequence $P_{n_k} \overset{w^*}{\to} P$. By closedness of $\mathscr{P}_A(X)$, we know that $P \in \mathscr{P}_A(X)$, hence $I(P, W_{Q_v}|R)$ is finite. This is a contradiction to $I(P_n, W_{Q_v}|R) \geq n$. Thus, the range of mutual information function is bounded.

Since the range of mutual information is bounded, it has a supremum. Let us denote this supremum value by $M$. By definition of supremum, for every $n$, there exists $P_n$ such that $I(P_n, W_{Q_v}|R) \geq M - \frac{1}{n}$. By weak\* compactness of $\mathscr{P}_A(X)$, there exists a weak\* convergent subsequence $P_{n_k} \overset{w^*}{\to} P$. By continuity of $I(P, W_{Q_v}|R)$, $\lim_k I(P_{n_k}, W_{Q_v}|R) \to I(P, W_{Q_v}|R)$. This requires that $M = I(P, W_{Q_v}|R)$ which means that the maximum is achieved by $P$. □

Since $\mathscr{P}_{g,\Gamma}(X)$ is weak\* compact and $I(P, W_{Q_v}|R)$ is continuous over $\mathscr{P}_{g,\Gamma}(X)$, by Proposition 4.1, there exists a capacity-achieving measure in $P_o \in \mathscr{P}_{g,\Gamma}(X)$. In the next subsection, we address a condition for the uniqueness of the capacity-achieving measure.

### 4.3 Uniqueness

In this subsection, we address sufficient conditions for the uniqueness of the capacity-achieving measure, a topic that that is of interest both from practical and theoretical standpoints.

**Proposition 4.2.** *Suppose $\mathscr{P}_A(X)$ is a convex set of input measures and $W_{Q_v}(\cdot|x)$ denotes a channel with side information. Assuming the existence of a capacity-achieving input measure $P_o$, it is unique upon the satisfaction of the hypothesis of Proposition 3.3.*

*Proof.* Suppose there exists another input measure $P_* \in \mathscr{P}_A(X)$ that achieves the capacity, also. For $P_o$ and $P_*$, if the hypothesis of Proposition 3.3 is satisfied, then their convex combination achieves a higher mutual information, which is a contradiction. □

## 4.4 Characterization

Now, we show how to characterize the capacity-achieving probability measure. Let $g: X \to \mathbb{R}^k$ be a continuous positive function that satisfies the hypothesis of Lemma 3.1, and let $G \in \mathbb{R}^{+k}$. By Lemma 3.1, the set of probability measures $\mathscr{P}_{g,\Gamma}(X)$ is weak* compact. Moreover, since the functionals $\int g_i dP$ are linear over the space of probability measures, the constraints $\int g_i dP \leq \Gamma_i$ make $\mathscr{P}_{g,\Gamma}(X)$ a convex set. Suppose that the mutual information function is weak* continuous over $\mathscr{P}_{g,\Gamma}(X)$. By Proposition 4.1, the mutual information function assume its maximum over $\mathscr{P}_{g,\Gamma}(X)$. The problem is how to characterize this measure.

To characterize the capacity-achieving measure, we use the global theory of constrained optimization [20] which uses Lagrange multipliers to facilitate the optimization problem. Applying the results of [20, p. 217], we obtain the following result.

**Lemma 4.2.** *Let $C = \sup_{\mathscr{P}_{g,\Gamma}(X)} I(P, W_{Q_v}|R)$. Then, there exists an element $\gamma \in \mathbb{R}^{+k}$ such that*

$$C = \sup \left\{ I(P, W_{Q_v}|R) - \sum_{i=1}^{k} \gamma_i \left( \int g_i dP - \Gamma_i \right) : \text{ for all } P \in \mathscr{P}_{g,\Gamma}(X) \right\}.$$

*Furthermore, this supremum is achieved by a probability measure $P^* \in \mathscr{P}_{g,\Gamma}(X)$ such that $\gamma_i \int g_i dP_o = \gamma_i \Gamma_i$ for $i = 1, \cdots, k$.*

*Proof.* It suffices to show that our optimization problem satisfies the hypothesis of [20, Theorem 1, p. 217]. Here, we have $\mathscr{P}_{g,\Gamma}(X)$ as the convex space we are optimizing over, $\int g_i dP$ as the convex constraint functions, and the mutual information is a concave function where its negative is our objective function over $\mathscr{P}_{g,\Gamma}(X)$. As we have discussed before, $\mathscr{P}_{g,\Gamma}(X)$ is a nonempty, weak* compact, and convex set. Since mutual information is weak* continuous over it, $C$ is finite. By Theorem 1 in [20, p. 217], we deduce that there exists $\gamma \geq 0$ that satisfies the hypothesis. This concludes the proof of the first assertion. Moreover, since mutual information achieves its maximum over $\mathscr{P}_{g,\Gamma}(X)$, the second assertion holds. □

To obtain the optimum probability measure in Lemma 4.2, we need some simplifying necessary and sufficient conditions which we define as follows. Let

$$f(P) \triangleq I(P, W_{Q_v}|R) - \sum_{i=1}^{k} \gamma_i \left( \int g_i dP - \Gamma_i \right). \tag{9}$$

It can be seen that for every $P \in \mathscr{P}_{g,\Gamma}(X)$, $f(P) < \infty$. This comes from the finiteness of both terms. Note that the second term is finite by definition of $\mathscr{P}_{g,\Gamma}(X)$, and the finiteness of the first term follows from Proposition 4.1. The weak* continuity and the concavity of (9) follows, similarly. By definition of *Gateaux differential* [20, p. 171], if for $\theta \in [0, 1]$, the limit

$$\delta f(P_o, P) \triangleq \lim_{\theta \downarrow 0} \frac{1}{\theta}[f(\theta P + (1-\theta)P_o) - f(P_o)]. \tag{10}$$

exists, then we call it the differentiation of $f$ at $P_o$ with increment of $P$. If (10) exists for all $P \in \mathscr{P}_{g,\Gamma}(X)$, we say that $f$ is differentiable at $P_o$. We state and prove the following theorem.

**Theorem 4.2.** *The supremum of $f$ is obtained by $P_o \in \mathscr{P}_{g,\Gamma}(X)$ if and only if $f$ is differentiable at $P_o$ and $\delta f(P_o, P) \leq 0$ for every $P \in \mathscr{P}_{g,\Gamma}(X)$.*

*Proof.* To prove the necessity, take any $P \in \mathscr{P}_{g,\Gamma}(X)$. For $0 \leq \theta \leq 1$, let $P_\theta = \theta P + (1-\theta)P_o$. By convexity of $\mathscr{P}_{g,\Gamma}(X)$, $P_\theta \in \mathscr{P}_\Gamma(X)$. Since $f$ attains its supremum on $P_o$, then $f(P_\theta) \leq f(P_o)$ which implies that $\frac{f(P_\theta) - f(P_o)}{\theta} \leq 0$. This implies that $\delta f(P_o, P) \leq 0$ upon its existence.

Moreover, since $f$ is a concave function with respect to $P$, we know that $\theta f(P) + (1-\theta)f(P_o) \leq f(P_\theta)$. This implies that

$$f(P) - f(P_o) \leq \frac{1}{\theta}[f(P_\theta) - f(P_o)].$$

Since both $f(P)$ and $f(P_o)$ are finite, $\frac{1}{\theta}[f(P_\theta) - f(P_o)]$ is bounded below for all values of $\theta$. Since $\theta \to 0$ implies that $P_\theta \xrightarrow{w^*} P_o$, by weak* continuity of $f$, we have $f(P_\theta) \to f(P_o)$. Moreover, $\frac{1}{\theta}[f(P_\theta) - f(P_o)]$ is bounded, then the existence of its *deleted limit* at $\theta = 0$ is immediate [14, p. 175]. Therefore, for all $P \in \mathscr{P}_{g,\Gamma}(X)$, $\delta f(P_o, P)$ exists and $\delta f(P_o, P) \leq 0$. This concludes the proof of the necessity.

To prove sufficiency, we proceed by contradiction. Suppose the assertion is not true. That is, there exists a probability measure $P^*$ such that $f(P^*) > f(P_o)$. By concavity of $f$, we would have

$$f(\theta P_o + (1-\theta)P^*) \geq \theta f(P_o) + (1-\theta)f(P^*) \geq f(P_o)$$

which creates a contradiction to non-positiveness of the differentiation. □

To characterize the capacity-achieving probability measure $P_o$, by Theorem 4.2, it suffices to check the sign of $\delta f(P_o, P)$ for all $P \in \mathscr{P}_{g,\Gamma}(X)$. Recalling the finiteness of the following terms, one can easily verify that

$$\delta f(P_o, P) = \iint [D(W_{Q_v}(\cdot|x)\|P_o W_{Q_v}) - \sum_{i=1}^{k} \gamma_i g_i(x)] dP dR$$

$$- \iint [D(W_{Q_v}(\cdot|x)\|P_o W_{Q_v}) - \sum_{i=1}^{k} \gamma_i g_i(x)] dP_o dR$$

Noting that $P_o$ is the capacity-achieving measure, by Theorem 4.2 this means that

$$\iint [D(W_{Q_v}(\cdot|x)\|P_o W_{Q_v}) - \sum_{i=1}^{k} \gamma_i g_i(x)] dP dR \leq C - \sum_{i=1}^{k} \gamma_i \Gamma_i \quad (11)$$

for all $P \in \mathscr{P}_\gamma(X)$. The following result simplifies this condition.

**Theorem 4.3 (Kuhn-Tucker conditions).** *The capacity-achieving measure is $P_o$ if and only if there exists $\gamma \geq 0$ such that*

$$\forall x \in X, \quad \int D(W_{Q_v}(\cdot|x)\|P_o W_{Q_v}) dR - \sum_{i=1}^{k} \gamma_i g_i(x) \leq C - \sum_{i=1}^{k} \gamma_i \Gamma_i \quad (12)$$

*where the equality holds for $P_o$-almost everywhere.*

*Proof.* The inverse part can be verified immediately from Theorem 4.2 and (11). For the direct part, since $P$ is arbitrary, we can take $P$ as dirac measures in different points, which results in the asserted inequality. By (11) and Theorem 4.2 we conclude the optimality of $P_o$. For the rest of the assertion, suppose that it is not true. That is, there exists a set $E \in \mathcal{B}_X$ such that $P_o(E) > 0$. Now taking the integration of LHS of (12) and decomposing the integration over $E$ and $E^c$, one can verify that this assumption leads to the inequality $C - \sum_{i=1}^{k} \gamma_i \Gamma_i < C - \sum_{i=1}^{k} \gamma_i \Gamma_i$ which is a contradiction. $\square$

Theorem 4.3 provides the necessary and sufficient conditions for the capacity-achieving measure in its most general form for continuous alphabet channels with side information at the receiver. Similar results are known for finite alphabet channels [1], [21] and [22]. For these channels, systematic algorithms are known to find the capacity-achieving measure [22]. In contrast, such algorithms are not known for continuous alphabet channels. However, one might be able to find the solution of Theorem 4.3 for special classes of channels.

Because of the importance of Theorem 4.3, let us rephrase the assertion of Theorem 4.3

more intuitively. For a given probability measure $P$ on $(X, \mathcal{B}_X)$, the *support* is defined as

$$\mathcal{S}_X(P) = \{x \in X | \forall \text{ open } U \in \mathcal{B}_X \text{ that contains } x, \, P(U) > 0\}.$$

The capacity-achieving measure is such that the equality in (12) occurs if and only if $x \in \mathcal{S}_X(P)$.

In an effort to characterize the support of the capacity-achieving measure, suppose $X = \mathbb{C}^n$ and let define $\rho : X \to \mathbb{R}$ as

$$\rho(x) \triangleq \iint [D(W_{Q_v}(\cdot|x) \| P_o W_{Q_v}) - \sum_{i=1}^{k} \gamma_i g_i(x)] dPdR + \sum_{i=1}^{k} \gamma_i \Gamma_i - C. \tag{13}$$

Let $Z = \mathbb{C}^{2n}$ and consider the extension $\rho : Z \to \mathbb{C}$ by replacing $\text{Re}(x_i) = z_i$ and $\text{Im}(x_i) = z_{n+i}$, corresponding to a natural embedding $\xi : X \to Z$. This means that $\rho(z)$ is real-valued for $z \in \mathcal{R}_\rho(Z)$, where $\mathcal{R}_\rho(Z)$ denotes the range of $\rho$. For every set $U \subseteq Z$, let $X_U = \xi^{-1}(U \cap \mathcal{R}_\rho(Z))$ denote the inverse image of $U$ under $\xi$. Using the properties of *analytic functions* [23], we state and prove the following proposition.

**Proposition 4.3.** *Let $\rho(z)$ be analytic on an open set $U \subseteq Z$, and let $X_U$ be the inverse image of $U$ under $\xi$. If $\mathcal{S}_X(P_o) \cap X_U$ has an interior point, then $X_U \subseteq \mathcal{S}_X(P_o)$.*

*Proof.* Suppose $\mathcal{S}_X(P_o) \cap X_U$ has an interior point, say for example $x_o$. Then, there exists an $\epsilon > 0$ and an open ball of radius $\epsilon$ centered at $x_o$, $B_\epsilon(x_o)$, such that $B_\epsilon(x_o) \subseteq \mathcal{S}_X(P_o)$. This means that the $\rho(x) = 0$ on $B_\epsilon(x_o)$, and consequently $\rho(z) = 0$ on $\xi(B_\epsilon(x_o)) \cap U$. Let $z_o = \xi(x_0)$. Since $\rho(z)$ is analytic on $z_o \in U$, there exists an open ball $B_r(z_o) \in U$ (for some $r > 0$) such that $\rho(z)$ can be represented as a Taylor series expansion on $B_r(z_o)$ [24]. Since $\rho(x) = 0$ on $B_\epsilon(x_o)$, the coefficients of the Taylor expansion are all zero. This implies that $\rho(z) = 0$ on $B_r(z_o)$. By Uniqueness Theorem [24, p. 12], [23], we conclude that $\rho(z) = 0$ on $U$. This means that $\rho(x) = 0$ on $X_U$ which implies that $X_U \subseteq \mathcal{S}_X(P_o)$. □

By Proposition 4.3, one can verify that if for some channel, the function $\rho(z)$ is analytic on $Z$, then either the support includes no interior point or it is equal to $X$.

This concludes our discussion on capacity-analysis of continuous alphabet channels with side information. In Part II of this two-part paper, we use this framework to study the capacity analysis problem for multiple antenna channels.

# 5    Conclusion

In this part, we established a general analytical framework for capacity analysis of continuous alphabet channels with side information (at the receiver). We studied the mutual information of these channels along with some of its analytical properties such as strict concavity and continuity. We established novel necessary and sufficient conditions for strict concavity and continuity of the mutual information in the weak* topology. We used these results and addressed issues regarding the existence, uniqueness, and the expression of capacity-achieving measure.

The results of this work can be used for capacity analysis of different classes of channels. Specifically, as will be shown in the Part II of this paper, these results are useful for capacity assessment of multiple antenna fading channels, fast or slow, Rician or Rayleigh, with partial or no CSI at the receiver, where the input probability measure could be subject to any combination of moment constraints.

# Appendix

# A    Preliminaries

In this appendix, we discuss some analytical notions and properties that are used throughout this paper. Some of these results are new while others are the review of the previous work, which we restate them here for the sake of completeness.

## A.1    Weak* topology

Let $(X, \mathcal{B}_X)$ be an LCH Borel-measurable space. The *weak* topology* is defined as follows. Let $C_0(X)$ denote the space of continuous functions from $X$ to $\mathbb{R}$ which vanish at infinity, i.e.,

$$C_0(X) = \{f : X \to \mathbb{R} |\ f \text{ is continuous and it vanishes at infinity}\}.$$

By the Riesz representation Theorem [13], the dual space of $C_0(X)$ is isomorphic to the space of Radon measures $\mathcal{M}(X)$ over the measurable space $(X, \mathcal{B}_X)$ [13]. To study the effect of an operation over $\mathscr{M}(X)$, there are different topologies that can be considered on $\mathscr{M}(X)$. The only crucial requirement is that the topology should be well behaved with respect to the operation of interest. In probability theory, where the objects of interest are the set of probability measures $\mathscr{P}(X) \subset \mathscr{M}(X)$, weak* topology is used which is the weakest topology on $\mathscr{M}(X)$ defined as follows. For each $f \in C_0(X)$, and every open set $G \subseteq \mathbb{R}$, let

$$U(f, G) \triangleq \left\{ \mu \in \mathscr{M}(X) \middle| \int f d\mu \in G \right\}.$$

The collection of all subsets $U(f, G) \subset \mathscr{M}(X)$ forms a basis for weak* topology on $\mathscr{M}(X)$. The collection of all subsets which are formed by any arbitrary union or finite intersections of the basis subsets form the weak* topology.

## A.2 Convergence

In weak* topology, the convergence phenomenon is called *weak\* convergence*[5] and defined as follows. A sequence of probability measures converges weakly*, denoted by $P_n \xrightarrow{w^*} P$ if and only if $\int f dP_n \to \int f dP$ for all $f \in C_0(X)$ [13]. Since our focus is on probability measures $\mathscr{P}(X) \subset \mathscr{M}(X)$, where all measures have unit norm, this is equivalent to saying that a sequence of probability measures converges weakly*, $P_n \xrightarrow{w^*} P$, if and only if $\int f dP_n \to \int f dP$ for $f \in C_b(X)$, where

$$C_b(X) = \{f : X \to \mathbb{R} | \; f \text{ is continuous and bounded}\}$$

denotes the set of all bounded continuous functions.

Given two measures $\nu$ and $\mu$ over $(X, \mathcal{B}_X)$, $\nu$ is said to be *absolutely continuous* with respect to $\mu$ denoted by $\nu \ll \mu$, if for every $E \in \mathcal{B}_X$ such that $\mu(E) = 0$, with $\nu(E) = 0$. By the Lebesgue-Radon-Nickodym theorem [13], there exists a $\mu$-integrable function $f$ such that for every $E \in \mathcal{B}_X$, $\nu(E) = \int_E f d\mu$. The function $f$ is unique $\mu$-almost everywhere ($\mu$-a.e.) and is called the density (Radon-Nikodym derivatives) of $\nu$ with respect to $\mu$, denoted by $f = \frac{d\nu}{d\mu}$. As an example of a sequence of probability measures which is weak* convergent, let us consider the following proposition.

**Proposition A.1.** *Let $(P_n)$ be a sequence of probability measures which are absolutely con-*

---

[5]In textbooks on probability theory, the term *vague* is used instead of weak*.

tinuous with respect to some measure $\mu$ (e.g. Lebesgue measure). For each $n$, let $f_n = \frac{dP_n}{d\mu}$ denote the density of $P_n$ with respect to $\mu$, and let $f$ be a function such that $f_n \to f$ $\mu$-a.e. and $\int f d\mu = 1$. Then, $P_n \xrightarrow{w^*} P$, where $P$ is the probability measure defined as $P(E) = \int_E f d\mu$ for every $E \in \mathcal{B}_X$. Moreover, for every $E \in \mathcal{B}_X$, $P(E) = \lim_n P_n(E)$.

*Proof.* Because $\{f_n\}$ are density functions for probability measures $\{P_n\}$ with respect to $\mu$, we have $\int f_n d\mu = 1$. By Fatou's lemma, for every $E \in \mathcal{B}_X$

$$P(E) = \int_E f d\mu \leq \liminf_n \int_E f_n d\mu = \liminf_n P_n(E).$$

By [17, p. 311], this implies the weak* convergence. Moreover, noting that $\int_E f d\mu + \int_{E^c} f d\mu = \int_E f_n d\mu + \int_{E^c} f d\mu = 1$, we deduce that

$$\int_E f d\mu = \lim_n \int_E f_n d\mu.$$

This concludes the second part of the assertion. □

To establish some of our results in this paper, it is of interest to verify whether the weak* convergence of a sequence of measures on one of these spaces implies the weak* convergence on the sequence of product measures. The following proposition is quite useful for this purpose.

**Proposition A.2.** *Let $(P_n)$ be a sequence of probability measures on $(X, \mathcal{B}_X)$ and let $T$ be a probability measure on $(Y, \mathcal{B}_Y)$. Then, $P_n \xrightarrow{w^*} P$ implies $(P_n \times T) \xrightarrow{w^*} (P \times T)$.*

*Proof.* For every open $E \in \mathcal{B}_X \otimes \mathcal{B}_Y$, let $E_y$ be as defined before. It is obvious that, for each $y$, $E_y$ is an open set in $\mathcal{B}_X$. Therefore,

$$\begin{aligned}
(P \times T)(E) &= \iint_E d(P \times T) \\
&= \int P(E_y) dT \quad \text{(By Tonelli's Theorem)} \\
&\leq \int \liminf_n P_n(E_y) dT \quad ([17, \text{p. } 311]) \\
&\leq \liminf_n \int P_n(E_y) dT \quad \text{(Fatou's lemma)} \\
&\leq \liminf_n (P_n \times T)(E) \quad \text{(By Tonelli's Theorem)}.
\end{aligned}$$

By [17, p. 311], this implies $(P_n \times T) \xrightarrow{w^*} (P \times T)$ and concludes the proof. □

Note that this can be also generalized for products of higher order. After this brief introduction to some necessary properties on convergence of probability measures, we now proceed to discuss the convergence of integrals, which is used to prove the continuity of mutual information.

## A.3 Uniform integrability

Some common sufficient conditions for convergence of a sequence of integrals are the monotone convergence theorem (MCT), the dominated convergence theorem (DCT), and the generalized dominated convergence theorem (GDCT) [13]. However, in this paper, we face a sequence of integrals whose convergence is not verifiable by any of these conditions. For our purposes, a less common condition exists known as *uniform integrability*.

Recalling that Radon probability measures are *regular* [13], i.e., for every $\epsilon > 0$, there exists a compact subset $K \in \mathcal{B}_X$ such that $P(K) \geq 1 - \epsilon$, we express the following definition.

**Definition A.1.** *Let $P \in \mathscr{P}(X)$. A collection of functions $\{f_\alpha\}_{\alpha \in A}$ is called uniformly P-integrable if*
$$\sup_{\alpha \in A} \int_{E_\alpha(c)} |f_\alpha| \, dP \to 0, \ as \ c \to \infty$$
*where $E_\alpha(c) = \{x \in X \mid |f_a| > c\}$.*

A more general definition of uniform integrability for positive measures is perhaps more familiar. However, we emphasize that Definition A.1 is an equivalent statement to the more general statement in the case of finite measures. We refer an interested reader for more details to [13, p. 92] and [17]. In the following theorem, we show that the sequence of integrals of a pointwise convergent sequence of uniformly P-integrable functions is converging.

**Theorem A.1.** *Let $P \in \mathscr{P}(X)$ and let $\{f_\alpha\}_{\alpha \in A}$ be uniformly P-integrable. Let $(f_n)$ be a sequence from $\{f_\alpha\}_{\alpha \in A}$ such that $f_n \to f$ P-almost everywhere (P-a.e.). Then, f is integrable, $\int f_n dP \to \int f dP$, and $\int |f_n - f| \, dP \to 0$.*

*Proof.* By definition of uniform integrability, for every $\epsilon \geq 0$,
$$\exists c_\epsilon \text{ such that } \forall \alpha \in A, \ \left| \int_{E_\alpha(c)} f_\alpha dP \right| \leq \frac{\epsilon}{3} \text{ for } c \geq c_\epsilon.$$

For every set $E$, let $\chi_E$ denote its characteristic function. Let $g_{n,c} = f_n \chi_{E_n^c(c)}$. Since $f_n \to f$ $P$-a.e., then $g_{n,c} \to g_c$ $P$-a.e. Because $|g_{n,c}(x)| \leq c$ for all $x$ and $n$, by DCT, we have $\int g_{n,c} dP \to \int g_c dP$. That is

$$\forall \epsilon > 0, \exists N \text{ such that } \left| \int_{E_n^c(c)} f_n dP - \int_{E^c(c)} f dP \right| \leq \frac{\epsilon}{3} \text{ for } n \geq N.$$

Now, by the triangular inequality

$$\left| \int f_n dP - \int f dP \right| \leq \left| \int_{E_n(c)} f_n dP \right| + \left| \int_{E_n^c(c)} f_n d\mu - \int_{E^c(c)} f d\mu \right| + \left| \int_{E(c)} f dP \right|$$
$$\leq \frac{\epsilon}{3} + \frac{\epsilon}{3} + \frac{\epsilon}{3} = \epsilon$$

This means that $\int f_n dP \to \int f dP$. To prove the other part of the assertion, we recall that since $|f_n - f| \leq |f_n| + |f|$, by GDCT it follows that $\int |f_n - f| dP \to 0$. $\square$

Another common scenario that arises in the context of convergence of integrals is the case that we have a fixed integrand function but a sequence of probability measures. To deal with such scenario, let us establish the following definition.

**Definition A.2.** *Let $\mathscr{P}_A(X)$ be a collection of probability measures over $(X, \mathcal{B}_X)$. A function $f$ is called uniformly integrable over $\mathscr{P}_A(X)$, if*

$$\sup_{P \in \mathscr{P}_A(X)} \int_{E(c)} |f| dP \to 0, \text{ as } c \to \infty$$

*where $E(c) = \{x \in X \mid |f| > c\}$.*

Using Definition A.2, we state and prove a sufficient condition for the convergence of the sequence of integrals of a function with respect to a weak* convergent sequence of probability measures.

**Theorem A.2.** *Let $\mathscr{P}_A(X)$ be a closed collection of probability measures and let $(P_n)$ be a weak* convergent sequence in it. If $f$ is a continuous function and uniformly integrable over $\{P_n\}$, then $\int f dP_n \to \int f dP$.*

*Proof.* For every $c > 0$, let $E(c) = \{x \in X \mid |f| > c\}$ and $\chi_{E(c)}$ be its characteristic function. By definition of uniform integrability of $f$ over $\{P_n\}$,

$$\forall \epsilon > 0, n, \exists c_\epsilon > 0 \text{ such that } \int_{E(c)} f dP_n \leq \frac{\epsilon}{3} \text{ for } c \geq c_\epsilon.$$

Let $g_c \triangleq f\chi_{E^c(c)} + c\chi_{E(c)}$. Continuity of $f$ over $X$ implies the continuity of $g_c$ over $X$. By weak* continuity of $\{P_n\}$,

$$\forall \epsilon > 0, \ \exists N \text{ such that } \left|\int g_c dP_n - \int g_c dP\right| \leq \frac{\epsilon}{3}, \text{ for } n \geq N.$$

By the triangular inequality,

$$\left|\int f dP_n - \int f dP\right| \leq \left|\int (f - g_c) dP_n\right| + \left|\int g_c dP_n - \int g_c dP\right| + \left|\int (f - g_c) dP\right|$$

$$\leq \left|\int_{E(c)} f dP_n\right| + \left|\int g_c dP_n - \int g_c dP\right| + \left|\int_{E(c)} f dP\right|$$

$$\leq \frac{\epsilon}{3} + \frac{\epsilon}{3} + \frac{\epsilon}{3} = \epsilon.$$

This means that $\int f dP_n \to \int f dP$ which concludes the proof. $\square$

This concludes our discussion on analytical preliminaries for the first part of this paper.